\documentclass[11pt,a4paper]{article}
\usepackage{amssymb}
\usepackage{amsmath}
\usepackage{amsthm}
\begin{document}
\title{\Large MacWilliams Theory over $\mathbb{Z}_k$ and nu-functions \\over Lattices}
\author{Zhiyong Zheng$^{1,2,3}$, Fengxia Liu$^{*2,1,3}$ and Kun Tian$^{*1,2,3}$
\\ \small{$^{1}$Engineering Research Center of Ministry of Education for Financial Computing}
\\ \small{and Digital Engineering, Renmin University of China, Beijing, 100872, China}
\\ \small{$^{2}$ Great Bay University, Great Bay Institute for  Advanced Study, Dongguan, 523800, China}
\\ \small{$^{3}$Henan Academy of Sciences, Zhengzhou, 450046, China}
\\ \small{$^{*}$Corresponding author email: shunliliu@gbu.edu.cn, tkun19891208@ruc.edu.cn}}

\date{}
\maketitle

\noindent\textbf{Abstract}\quad Continuing previous works on MacWilliams theory over codes and lattices, a generalization of the MacWilliams theory over $\mathbb{Z}_k$ for $m$ codes is established, and the complete weight enumerator MacWilliams identity also holds for codes over the finitely generated rings $\mathbb{Z}_k[\xi]$. In the context of lattices, the analogy of the MacWilliams identity associated with nu-function was conjectured by Sol\'{e} in 1995, and we present a new formula for nu-function over the lattices associated with a ternary code, which is rather different from the original conjecture. Furthermore, we provide many counterexamples to show that the Sol\'{e} conjecture never holds in the general case, except for the lattices associated with a binary code.

\noindent \textbf{Keywords:} MacWilliams theorem, codes, lattices, nu-function.

\section{Introduction}

The MacWilliams theorem for linear codes with a Hamming metric \cite{11,12} establishes an identity that relates the weight enumerator of a code to that of its dual code. Various authors have extended this study in different directions. One direction involves generalizing the weight enumerators to include more than two variables, such as Lee and complete weight enumerators, and extending the concept to codes defined over alphabets beyond finite fields. For instance, Wan \cite{21} provided the MacWilliams theorem for code over Galois rings. Another generalization involves adapting the notion of weight to consider multiple codewords simultaneously, leading to the generalized Hamming weights by Wei \cite{22} and MacWilliams-type results for $m$-tuple support enumerators by Kl{\o}ve \cite{09}, Shiromoto \cite{18}, Simonis \cite{19}, and Ray-Chaudhuri and Siap \cite{14,15}. Britz \cite{04,05} further generalized some of these results and provided matroid-theoretic proofs. Britz \cite{03} outlined new and extensive connections between weight enumerators and Tutte polynomials of matroids.

 First, we consider the MacWilliams theorem for $m$ codes over $\mathbb{Z}_k$, where the ring of integers modulo $k\ (k>1)$. We generalize the results of \cite{002} and establish a MacWilliams-type identity for $m$-tuple support enumerator over the finite ring $\mathbb{Z}_k$. We prove that the MacWilliams theorem of the complete weight enumerators also holds for code over $\mathbb{Z}_k[\xi]$. For MacWilliams theorems over more general finite rings, for example, Frobenius rings and finite modules, we refer to Wood \cite{006} and \cite{007}.

 The close relationships between codes and lattices have been intensively studied over the past few decades. The classical problem of counting lattice points in Euclidean spheres involves the use of the Jacobian theta function. The MacWilliams formula can be viewed as a finite counterpart to the Poisson summation formula employed in theta function theory within this framework. The nu-function of lattices is involved in the more contemporary issue of enumerating lattice points in pyramids for the $L^1$-norm \cite{02,20}. These two challenges have important applications in the field of multidimensional vector quantization \cite{02,17}. In this paper, we present a new analog of the MacWilliams identity over lattices involving a nut function, a problem conjectured by Sol\'{e} \cite{25,20}. Here, we provide a solution to this problem.

\subsection{$m$-Tuple MacWilliams Identity}

To begin, we present the essential definitions required to articulate MacWilliams's original theorem \cite{12}. Consider $\mathbb{F}_q$, a finite field containing $q$ elements, $n$ as a positive integer, and $C\subset \mathbb{F}_q^n$ representing a linear code. Let $|C|$ denote the total number of codewords in $C$, and $<a,b>$ represent the standard pairing on $\mathbb{F}_q^n$. For any $x\in \mathbb{F}_q^n$, the Hamming weight, symbolized as $w(x)$, is defined as the count of non-zero coordinates in $x$. We use $W_C(z)$ to signify the Hamming weight enumerator of $C$ in the variables $z$:
\begin{equation*}
W_C(z)=\sum\limits_{x\in C} z^{w(x)},
\end{equation*}
or an equivalent homogeneous form in two indeterminates $z_1$ and $z_2$
\begin{equation*}
W_C(z_1,z_2)=\sum\limits_{x\in C} z_1^{w(x)}z_2^{n-w(x)}.
\end{equation*}
Let $C\subset \mathbb{F}_q^n$ be any linear code and $C^{\bot}$ be its dual code, MacWilliams \cite{07} (or also see \cite{11}) showed that
\begin{equation*}
W_{C^{\bot}}(z)=\frac{1}{|C|}(1+(q-1)z)^n W_C \left(\frac{1-z}{1+(q-1)z}\right).  \tag{1.1}
\end{equation*}
Let $z=z_1/z_2$, one has the homogeneous form of MacWilliams identity immediately (see Theorem 9.32 of \cite{10})
\begin{equation*}
W_{C^{\bot}}(z_1,z_2)=\frac{1}{|C|} W_C(z_2-z_1,z_2+(q-1)z_1).  \tag{1.2}
\end{equation*}

 Klemm \cite{002} expanded upon the MacWilliams identity in 1987, applying it to Hamming weight enumerators for codes defined over finite rings, $\mathbb{Z}_k$. To explain this concept, consider $\mathbb{Z}_k$ as the ring of integers modulo $k$ (where $k>1$) and a code $C\subset \mathbb{Z}_k^n$ as an additive subgroup of $\mathbb{Z}_k^n$. For any codeword $x\in \mathbb{Z}_k^n$, the Hamming weight $w(x)$ is defined in the conventional manner. This finding was derived from\cite{002}:

\textbf{Theorem 1}\ \ (Klemm) Let $C\subset \mathbb{Z}_k^n$ be a code of length $n$, $C^{\bot}$ be its dual, then
\begin{equation*}
W_{C^{\bot}}(z)=\frac{1}{|C|}(1+(k-1)z)^n W_C \left(\frac{1-z}{1+(k-1)z}\right)  \tag{1.3}
\end{equation*}
or equivalently
\begin{equation*}
W_{C^{\bot}}(z_1,z_2)=\frac{1}{|C|} W_C(z_2+(k-1)z_1,z_2-z_1).  \tag{1.4}
\end{equation*}

  A primary objective of this research is to expand upon Theorem 1 and introduce a MacWilliams-type identity for the $m$-tuple support enumerator defined over $\mathbb{Z}_k$. To present our results, consider $m$ codes $C_1,C_2,\cdots,C_m \subset \mathbb{Z}_k^n$, which may or may not be identical. Let $\mathbb{Z}_k^{m\times n}$ represent the ring of all $m\times n$ matrices over $\mathbb{Z}_k$. For a matrix $x\in \mathbb{Z}_k^{m\times n}$, we denote $ew(x)$ as the effective length weight of $x$, which is the count of non-zero columns in $x$. Define $\underline{C}=C_1\times C_2\times\cdots \times C_m$, $\underline{C}^{\bot}=C_1^{\bot}\times C_2^{\bot}\times \cdots \times C_m^{\bot}$, and $|\underline{C}|=|C_1|\times \cdots \times |C_m|$. We consider any element $x=(x_1,x_2,\cdots,x_m)\in \underline{C}$ as a matrix in $\mathbb{Z}_k^{m\times n}$ by
\begin{equation*}
x=\begin{pmatrix} x_1 \\ x_2 \\ \vdots \\ x_m \end{pmatrix} \in \mathbb{Z}_k^{m\times n},\quad x_i\in C_i.
\end{equation*}

\begin{equation*}
W_{\underline{C}}^{(m)}(z)=\sum\limits_{x\in \underline{C}} z^{ew(x)}.
\end{equation*}

\textbf{Theorem 2}\ \ Let $C_1,C_2,\cdots,C_m$ be any $m$ linear codes over $\mathbb{Z}_k^n$, then we have
\begin{equation*}
W_{{\underline{C}}^{\bot}}^{(m)}(z)=\frac{1}{|\underline{C}|} (1+(k^m-1)z)^n W_{\underline{C}}^{(m)}\left(\frac{1-z}{1+(k^m-1)z}\right).  \tag{1.5}
\end{equation*}

 MacWilliams theorems have been extended in various ways, with a significant number of these generalizations addressing complete weight enumerators, as exemplified by the work of Wan \cite{21} and Kaplan \cite{08}. In our effort to further expand upon Theorem 1, we investigate a code defined over the finitely generated ring $\mathbb{Z}_k^n[\xi]$. Our research establishes that the MacWilliams theorem for complete weight enumerators extends its applicability to codes over $\mathbb{Z}_k^n[\xi]$.

Let $R=\mathbb{Z}_k[\xi]$ be a finitely generated ring with $k$ being a positive integer, $\xi$ be the root of an irreducible polynomial of degree $t$, $C\subset R^n$ be a linear code. For any $x=x_1 x_2\cdots x_n\in C$, assuming $x_i=x_{i0}+x_{i1}\xi+\cdots+x_{i(t-1)}\xi^{t-1}\in R$, we define the function $u(x_i)=x_{i0}+x_{i1}\cdot k+\cdots+x_{i(t-1)}\cdot k^{t-1}\in [0,k^t-1]$, and $\tau(x_i)=(x_{i0},x_{i1},\cdots,x_{i(t-1)})\in \mathbb{Z}_k^t$. For $j=0,1,\cdots,k^{t}-1$, let $n_j(x)=\#\{1\leqslant i\leqslant n\ |\ u(x_i)=j\}$, which is the number of $x_i$ that satisfies $u(x_i)=j$.  We define the complete weight enumerator of $C$ as the following
\begin{equation*}
W_C(z_0,z_1,\cdots,z_{|R|-1})=\sum\limits_{x\in C} z_0^{n_0(x)} z_1^{n_1(x)} \cdots z_{|R|-1}^{n_{|R|-1}(x)}.
\end{equation*}
We have the following MacWilliams identity.

\textbf{Theorem 3}\ \ Let $C\subset \mathbb{R}^n$ be a linear code of length $n$, $C^{\bot}$ be its dual, then
\begin{equation*}
W_{C^{\bot}}(z_0,z_1,\cdots,z_{|R|-1})=
\end{equation*}
\begin{equation*}
\frac{1}{|C|} W_C \left(\sum\limits_{l=0}^{|R|-1}z_l,\sum\limits_{l=0}^{|R|-1}e^{\frac{2\pi <\tau(1),\tau(l)>}{k}i}z_l,\cdots,\sum\limits_{l=0}^{|R|-1}e^{\frac{2\pi<\tau(|R|-1),\tau(l)>}{k}i}z_l\right),  \tag{1.6}
\end{equation*}
where $\tau(l)$ is the $k$ base vector of $l$ with length $t$, i.e. if $l=l_0+l_1\cdot k+\cdots+l_{t-1}\cdot k^{t-1}$, $l_0,l_1,\cdots,l_{t-1}\in \mathbb{Z}_{k}$, then $\tau(l)=(l_0,l_1,\cdots,l_{t-1})$.

Wan \cite{21} presented a specific case of (1.6) in his work on Galois rings. In this paper, Theorem 3 represents a more generalized form of the main conclusion of \cite{21}.

In the finite ring $\mathbb{Z}_k$, it is noteworthy that researchers are increasingly focusing on the weight enumerator with the Lee metric. For further information, refer to the results in \cite{001,003,004,005}. Notably, \cite{001} demonstrated that the analogy of the MacWilliams theorem with Lee weight enumerators does not exist for $\mathbb{Z}_k$ when $k\geqslant 5$. For MacWilliams theorems over Frobenius rings, refer to Wood \cite{006} and \cite{007}.

\subsection{Nu-Function over Lattices}

Let $\Lambda\subset \mathbb{R}^n$ be a lattice of rank $n$, the nu-function over $\Lambda$ is defined by
 \begin{equation*}
 \nu_{\Lambda}(z)=\sum\limits_{x\in \Lambda}z^{|x|_1}=\sum\limits_{n=0}^{+\infty} z^n |\{x\in \Lambda: |x|_1=n\}|,
 \end{equation*}
 where $|x|_1=\sum\limits_{i=1}^n |x_i|$ is the $L^1$-norm. Although the Poisson summation formula for the nu-function is unknown, the same analogy of the MacWilliams identity exists. It was conjectured by \cite{20} that
 \begin{equation*}
  2^{\frac{n}{2}} \nu_{\Lambda^*}(\text{tanh}^2 (\frac{\beta}{2}))=\text{det}(\Lambda)(\text{sinh}(2\beta))^{\frac{n}{2}}\nu_{\Lambda}(\text{tanh}(\frac{\alpha}{2})),  \tag{1.7}
  \end{equation*}
    where $\Lambda$ is the arbitrary lattice of rank $n$, and the parameters $\alpha$ and $\beta$ are connected by the relation $e^{-2\beta}=\text{tanh}(\alpha)$. In \cite{008}, we demonstrated that the aforementioned conjecture holds true for a lattice associated with a binary code $C$. However, in general, the following result differs significantly from the aforementioned conjecture. Initially, we present a novel formula for the nu-function associated with a ternary code, and subsequently, we provide numerous counter-examples to elucidate why the conjecture is invalid in general cases. To articulate our primary result, we first generalize the construction $A$ as follows:

\textbf{Construction $A_k$}

Let $C\subset \mathbb{Z}_k^n$ be a code. A lattice $A_k(C)$ is constructed by
\begin{equation*}
A_k(C)=\{x=(x_1,\cdots,x_n)\in \mathbb{Z}^n\ \big|\ \exists c\in C\ \text{such that}\ x\equiv c\ (\text{mod}\ k)\}.
\end{equation*}

\textbf{Theorem 4}\ \ If $\Lambda=A_3(C)$ is a lattice associated with a ternary code $C$, then we have
\begin{equation*}
3^n \nu_{\Lambda^*}(\text{tanh}^3 (\frac{\beta}{2}))=\text{det}(\Lambda)\left[\frac{(1+\text{tanh}(\frac{\beta}{2}))(1-\text{tanh}^3(\frac{\alpha}{2}))}{(1-\text{tanh}(\frac{\beta}{2}))(1+\text{tanh}^3(\frac{\alpha}{2}))}\right]^n \nu_{\Lambda}(\text{tanh}(\frac{\alpha}{2})),  \tag{1.8}
\end{equation*}
where the parameters $\alpha$ and $\beta$ are connected by the relation
\begin{equation*}
e^{-2\beta}=\frac{3\text{tanh}(\alpha)}{8-5\text{tanh}(\alpha)}.
\end{equation*}

 Upon comparing equations (1.8) and (1.7), it is evident that equation (1.8) does not hold for the general lattice, with the exception of $\Lambda=A_2(C)$, as demonstrated in \cite{008}. For lattices $\Lambda=A_k(C)\ (k>3)$, numerous counter-examples are provided in the final section.

\textbf{Property}\ \ The relationship between parameters $\alpha$ and $\beta$ satisfying the following symmetrical relation
\begin{equation*}
e^{-2\beta}=\frac{3\text{tanh}(\alpha)}{8-5\text{tanh}(\alpha)}\Longleftrightarrow e^{-2\alpha}=\frac{3\text{tanh}(\beta)}{8-5\text{tanh}(\beta)}.
\end{equation*}

\textbf{Proof:} We have
\begin{equation*}
e^{-2\beta}=\frac{3\text{tanh}(\alpha)}{8-5\text{tanh}(\alpha)}\Longleftrightarrow e^{-2\beta}=\frac{3\cdot\frac{e^{\alpha}-e^{-\alpha}}{e^{\alpha}+e^{-\alpha}}}{8-5\cdot\frac{e^{\alpha}-e^{-\alpha}}{e^{\alpha}+e^{-\alpha}}}=\frac{3e^{2\alpha}-3}{3e^{2\alpha}+13}
\end{equation*}
\begin{equation*}
\Longleftrightarrow e^{2\alpha}=\frac{3+13e^{-2\beta}}{3-3e^{-2\beta}} \Longleftrightarrow e^{-2\alpha}=\frac{3e^{2\beta}-3}{3e^{2\beta}+13}=\frac{3\text{tanh}(\beta)}{8-5\text{tanh}(\beta)}.
\end{equation*}
\hspace*{12cm} $\Box$

\section{Finite Fourier Transform over the Matrix Ring $\mathbb{Z}_k^{m\times n}$}

Let $\mathbb{Z}_k$ be the ring of integer modulo $k$ with $k>1$, $e^{2\pi i/k}$ be the primitive root of the unit, and an additive characteristic $\psi$ for $\mathbb{Z}_k$ by $\psi(a)=e^{2\pi ia/k}\ (a\in \mathbb{Z}_k)$. Suppose that $H\subset \mathbb{Z}_k$ and $H\neq 0$, is an additive subgroup of $\mathbb{Z}_k$, trivially, we have
\begin{equation*}
\sum\limits_{a\in H}\psi(a)=\sum\limits_{a\in H}e^{2\pi ia/k}=\left\{\begin{array}{cc} 0,&\ \text{if}\ a\neq 0,\\ |H|,&\ \text{if}\ a=0. \end{array}\right.  \tag{2.1}
\end{equation*}
If $f(x)$ is a function defined over $\mathbb{Z}_k^n$, the finite Fourier transform of $f$ is defined by
\begin{equation*}
\text{FT}f(x)=\sum\limits_{\xi\in \mathbb{Z}_k^n} f(\xi)\psi(<x,\xi>),\quad \forall x\in \mathbb{Z}_k^n,
\end{equation*}
where $<x,\xi>$ is the typical pairing of $x$ and $\xi$.

Let $C\subset \mathbb{Z}_k^n$ be a linear code of length $n$ and $C^{\bot}$ be its dual code. We define its characteristic function as $\chi_C(x)$, that is, $\chi_C(x)=1$, if $x\in C$, and $\chi_C(x)=0$, if $x\notin C$.

\textbf{Lemma 2.1}\ \ Suppose that $C\subset \mathbb{Z}_k^n$ is a linear code, and $C^{\bot}$ is its dual, then we have
\begin{equation*}
\chi_{C^{\bot}}(x)=\frac{1}{|C|}\text{FT}\chi_{C}(x).
\end{equation*}

\textbf{Proof:} By the definition, we only need to show that
\begin{equation*}
\sum\limits_{\xi\in C}\psi(<x,\xi>)=\left\{\begin{array}{cc} |C|,&\ \text{if}\ x\in C^{\bot}.\\ 0,&\ \text{if}\ x\notin C^{\bot}. \end{array}\right.
\end{equation*}
If $x\notin C^{\bot}$, let $T(x)$ be the subgroup of $C$,  given by
\begin{equation*}
T(x)=\{y\in C\ \big|\ <y,x>=0\}.
\end{equation*}
 It is easily seen that $<y_1,x>=<y_2,x>$ if $y_1-y_2\in T(x)$, it means that $<y_1,x>=<y_2,x>$ if $y_1$ and $y_2$ lie in the same coset of $T(x)$. Thus, by  (2.1), we
\begin{equation*}
\sum\limits_{\xi\in C}\psi(<x,\xi>)=|T(x)|\sum\limits_{c\in C/T(x)} \psi(<c,x>)=|T(x)|\sum\limits_{a\in H} \psi(a)=0,
\end{equation*}
where $H$ is a subgroup given by
\begin{equation*}
H=\{<c,x>\ \big|\ c\in C/T(x)\}.
\end{equation*}
\hspace*{12cm} $\Box$

From Lemma 2.1, we trivially have the Poisson summation formula for any function $f$ over $\mathbb{Z}_k^n$
\begin{equation*}
\sum\limits_{x\in C^{\bot}}f(x)=\frac{1}{|C|}\sum\limits_{x\in C}\text{FT}f(x).  \tag{2.2}
\end{equation*}
If taking $f(x)=z^{w(x)}$, it is easy to see that
\begin{equation*}
\text{FT}z^{w(x)}=(1+(k-1)z)^n \left(\frac{1-z}{1+(k-1)z}\right)^{w(x)}.
\end{equation*}
Then, Theorem 1 follows immediately.

To prove Theorem 2, we shall first generalize this transform to a matrix ring. Let $\mathbb{Z}_k^{m\times n}$ be a ring comprising all $m\times n$ matrices over $\mathbb{Z}_k$. For any two matrices $x,y\in \mathbb{Z}_k^{m\times n}$, $x^T y$ is an $n\times n$ square matrix. Let Tr$(x^T y)$ denote the trace of $x^T y$, then Tr$(x^T y)\in \mathbb{Z}_k$. We define $<x,y>=\text{Tr}(x^T y)$ as the pairing of $x$ and $y$.

\textbf{Definition 2.1}\ \ Suppose that $f$ is a function defined over $\mathbb{Z}_k^{m\times n}$, the finite Fourier transform FT$f$ of $f$ is defined by
\begin{equation*}
\text{FT}f(x)=\sum\limits_{\xi\in \mathbb{Z}_k^{m\times n}} f(\xi)\psi(<x,\xi>),\quad x\in \mathbb{Z}_k^{m\times n}.
\end{equation*}
It is easy to see that
\begin{equation*}
f(x)=\frac{1}{q^{mn}} \sum\limits_{\xi\in \mathbb{Z}_k^{m\times n}} \text{FT}f(\xi) \psi(-<x,\xi>),\quad x\in \mathbb{Z}_k^{m\times n}.
\end{equation*}
Let $C_1,C_2,\cdots,C_m$ be $m$ codes over $\mathbb{Z}_k$ of length $n$, $\underline{C}=C_1\times \cdots \times C_m$. We consider $\underline{C}$ as a subset of $\mathbb{Z}_k^{m\times n}$ according to $x=(x_1,\cdots,x_m)^T$, where $x_i\in C_i$. Let $\chi_{\underline{C}}$ denote the characteristic function of $\underline{C}$, such that $\chi_{\underline{C}}(x)=1$ if $x\in \underline{C}$, and $\chi_{\underline{C}}(x)=0$ if $x\notin \underline{C}$.

\textbf{Lemma 2.2}\ \ If $C_i$ ($1\leqslant i\leqslant m$) is the linear code of $\mathbb{Z}_k^n$, then we have
\begin{equation*}
\chi_{\underline{C}^{\bot}}(x)=\frac{1}{|\underline{C}|}\text{FT}\chi_{\underline{C}}(x),
\end{equation*}
where $\underline{C}^{\bot}=C_1^{\bot}\times C_2^{\bot}\times\cdots \times C_m^{\bot}$.

\textbf{Proof:} By the definition of FT$\chi_{\underline{C}}$, we observe  that for any $x\in \mathbb{Z}_k^{m\times n}$,
\begin{equation*}
\text{FT}\chi_{\underline{C}}(x)=\sum\limits_{\xi\in \mathbb{Z}_k^{m\times n}} \chi_{\underline{C}}(\xi)\psi(<x,\xi>)=\sum\limits_{\xi\in \underline{C}}\psi(<x,\xi>).
\end{equation*}
If we write
\begin{equation*}
x=\begin{pmatrix} x_1\\ \vdots\\ x_m \end{pmatrix},\ x_i\in \mathbb{Z}_k^n,\ \text{and}\ \xi=\begin{pmatrix} \xi_1\\ \vdots\\ \xi_m \end{pmatrix},\ \xi_i\in C_i,
\end{equation*}
then
\begin{equation*}
<x,\xi>=\text{Tr}(x^T \xi)=\sum\limits_{i=1}^m <x_i,\xi_i>.
\end{equation*}
It follows that
\begin{equation*}
\text{FT}\chi_{\underline{C}}(x)=\mathop{\prod}\limits_{i=1}^m \sum\limits_{\xi_i\in C_i} \psi(<x_i,\xi_i>).
\end{equation*}
It is known that
\begin{equation*}
\sum\limits_{\xi_i\in C_i} \psi(<x_i,\xi_i>)=\left\{\begin{array}{cc} |C_i|, & \text{if}\ x_i\in C_i^{\bot}. \\ 0, & \text{otherwise.} \end{array}\right.
\end{equation*}
We have FT$\chi_{\underline{C}}(x)=|\underline{C}|\chi_{\underline{C}^{\bot}}(x)$.\\
\hspace*{12cm} $\Box$

Based on the aforementioned observation, we derive the following more generalized Poisson summation formula applicable to the matrix ring:

\textbf{Lemma 2.3}\ \ Let $\underline{C}\subset \mathbb{Z}_k^{m\times n}$ be the block linear codes over $\mathbb{Z}_k$, and $f(x)$ be arbitrary function over $\mathbb{Z}_k^{m\times n}$, then
\begin{equation*}
\sum\limits_{x\in \underline{C}^{\bot}}f(x)=\frac{1}{|\underline{C}|} \sum\limits_{x\in \underline{C}}\text{FT}f(x).
\end{equation*}

\textbf{Proof:}
\begin{equation*}
\sum\limits_{x\in \underline{C}^{\bot}}f(x)=\sum\limits_{x\in \mathbb{Z}_k^{m\times n}} \chi_{\underline{C}^{\bot}}(x)f(x)
\end{equation*}
\begin{equation*}
\qquad\qquad\qquad\quad=\frac{1}{|\underline{C}|} \sum\limits_{x\in \mathbb{Z}_k^{m\times n}} f(x)\text{FT}\chi_{\underline{C}}(x)
\end{equation*}
\begin{equation*}
\qquad\qquad\qquad\qquad\qquad\qquad\qquad=\frac{1}{|\underline{C}|} \sum\limits_{\xi\in \mathbb{Z}_k^{m\times n}} \chi_{\underline{C}}(\xi) \sum\limits_{x\in \mathbb{Z}_k^{m\times n}} f(x)\psi(<x,\xi>)
\end{equation*}
\begin{equation*}
\qquad\quad=\frac{1}{|\underline{C}|} \sum\limits_{\xi\in \underline{C}}\text{FT}f(\xi).
\end{equation*}
\hspace*{12cm} $\Box$

Subsequently, we present a concise proof for Theorem 2.

\textbf{Proof of Theorem 2:}

Let $f(x)=z^{ew(x)}$, where $x\in \mathbb{Z}_k^{m\times n}$ and $ew(x)$ are the effective length weight functions given in Section 1. To prove Theorem 2, using the Poisson summation formula, we show only that
\begin{equation*}
\text{FT}z^{ew(x)}=(1+(k^m-1)z)^n \left(\frac{1-z}{1+(k^m-1)z}\right)^{ew(x)}.
\end{equation*}

 Write $x=(x_1,x_2,\cdots,x_n)\in \mathbb{Z}_k^{m\times n}$, $x_i\in \mathbb{Z}_k^m$, and $\xi=(\xi_1,\xi_2,\cdots,\xi_n)$ $\in \mathbb{Z}_k^{m\times n}$, $\xi_i\in \mathbb{Z}_k^m$, then
\begin{equation*}
<x,\xi>=\text{Tr}(x^T \xi)=\sum\limits_{i=1}^n <\xi_i,x_i>.
\end{equation*}
It follows that
\begin{equation*}
\text{FT}f(x)=\sum\limits_{\xi\in \mathbb{Z}_k^{m\times n}} f(\xi) \psi(<x,\xi>)
\end{equation*}
\begin{equation*}
\qquad\qquad\ =\sum\limits_{\xi\in \mathbb{Z}_k^{m\times n}} z^{ew(\xi)} \psi(<x,\xi>)
\end{equation*}
\begin{equation*}
\qquad\qquad\qquad\qquad\ \ \ =\sum\limits_{\xi\in \mathbb{Z}_k^{m\times n}} z^{ew(\xi_1)+\cdots+ew(\xi_n)} \psi(<x,\xi>)
\end{equation*}
where $ew(\xi_i)=1$, if $\xi_i\neq 0$, and $ew(\xi_i)=0$, if $\xi_i=0$ is a zero vector. Therefore, we have
\begin{equation*}
\text{FT}f(x)=\mathop{\prod}\limits_{i=1}^n (\sum\limits_{\xi_i\in \mathbb{Z}_k^m} z^{ew(\xi_i)} \psi(<x_i,\xi_i>))
\end{equation*}
\begin{equation*}
\qquad\qquad\quad=\mathop{\prod}\limits_{i=1}^n (1+z\sum\limits_{\xi_i\in \mathbb{Z}_k^m \backslash \{0\}} \psi(<x_i,\xi_i>)).
\end{equation*}
If $x_i=0$, it is easy to obtain
\begin{equation*}
\sum\limits_{\xi_i\in \mathbb{Z}_k^m \backslash \{0\}} \psi(<x_i,\xi_i>)=k^m-1,
\end{equation*}
otherwise, if $x_i\neq 0$, we write $x_i=(x_{i1},\cdots,x_{im})'$ and $\xi_i=(\xi_{i1},\cdots,\xi_{im})'$, then $\exists 1\leqslant s\leqslant m$, such that $x_{is}\neq 0$. Based on (2.1), we know that
\begin{equation*}
\sum\limits_{\xi_{is}\in \mathbb{Z}_k} \psi(x_{is}\xi_{is})=0,
\end{equation*}
thus,
\begin{equation*}
\sum\limits_{\xi_i\in \mathbb{Z}_k^m} \psi(<x_i,\xi_i>)=\mathop{\prod}\limits_{j=1}^m \sum\limits_{\xi_{ij}\in \mathbb{Z}_k} \psi(x_{ij}\xi_{ij})=0.
\end{equation*}
So we get
\begin{equation*}
\sum\limits_{\xi_i\in \mathbb{Z}_k^m \backslash \{0\}} \psi(<x_i,\xi_i>)=\left\{\begin{array}{cc} k^m-1, & \text{if}\ x_i=0. \\ -1, & \text{if}\ x_i\neq 0. \end{array}\right.
\end{equation*}
It follows that
\begin{equation*}
\text{FT}z^{ew(x)}=\mathop{\prod}\limits_{i=1}^n (1+z\sum\limits_{\xi_i\in \mathbb{Z}_k^m \backslash \{0\}} \psi(<x_i,\xi_i>))\qquad
\end{equation*}
\begin{equation*}
\qquad\qquad\quad\ =\mathop{\prod}\limits_{i=1}^n ((1-z)^{ew(x_i)}(1+(k^m-1)z)^{1-ew(x_i)})
\end{equation*}
\begin{equation*}
\qquad\qquad\ =(1+(k^m-1)z)^n \left(\frac{1-z}{1+(k^m-1)z}\right)^{ew(x)}.
\end{equation*}
We complete the proof of Theorem 2.\\
\hspace*{12cm} $\Box$

Finally, for the finitely generated ring $R=\mathbb{Z}_k[\xi]$, where $k$ is a positive integer and $\xi$ is a root of an irreducible polynomial of degree $t$, and linear code $C\subset R^n$, we consider the function
 \begin{equation*}
  f(x)=z_0^{n_0(x)} z_1^{n_1(x)} \cdots z_{|R|-1}^{n_{|R|-1}(x)}.
  \end{equation*}
Employing a methodology analogous to that of Theorem 2, we proceed to demonstrate the validity of Theorem 3.

\textbf{Proof of Theorem 3:} The main problem is the calculation of the finite Fourier transform $f(x)=z_0^{n_0(x)} z_1^{n_1(x)} \cdots z_{|R|-1}^{n_{|R|-1}(x)}$. We write $x=x_1 x_2\cdots x_n$ $\in C\subset R^n$, and $y=y_1 y_2\cdots y_n\in R^n$, and $<x,y>=\sum\limits_{j=1}^n <\tau(x_j),\tau(y_j)>$, here $\tau(x_j)$, $\tau(y_j)$ are the coefficient vectors of $x_j$ and $y_j$, then
\begin{equation*}
\text{FT}f(x)=\sum\limits_{y\in R^n} f(y)\psi(<x,y>)
\end{equation*}
\begin{equation*}
\hspace{-2.2cm}=\sum\limits_{y\in R^n} z_0^{n_0(y)} z_1^{n_1(y)} \cdots z_{|R|-1}^{n_{|R|-1}(y)}\psi(<x,y>)
\end{equation*}
\begin{equation*}
=\mathop{\prod}\limits_{j=1}^n \sum\limits_{y_j\in R} z_0^{n_0(y_j)} z_1^{n_1(y_j)} \cdots z_{|R|-1}^{n_{|R|-1}(y_j)}\psi(<\tau(x_j),\tau(y_j)>)
\end{equation*}
\begin{equation*}
\hspace{-5cm}=\mathop{\prod}\limits_{j=1}^n \sum\limits_{l=0}^{|R|-1} e^{\frac{2\pi <\tau(x_j),\tau(l)>}{k}i} z_l
\end{equation*}
\begin{equation*}
\hspace{-0.5cm}=\left(\sum\limits_{l=0}^{|R|-1} z_l\right)^{n_0(x)} \left(\sum\limits_{l=0}^{|R|-1} e^{\frac{2\pi <\tau(1),\tau(l)>}{k}i} z_l\right)^{n_1(x)} \cdots \left(\sum\limits_{l=0}^{|R|-1} e^{\frac{2\pi <\tau(|R|-1),\tau(l)>}{k}i} z_l\right)^{n_{|R|-1}(x)}.
\end{equation*}
Based on the Poisson summation formula, we have Theorem 3 immediately.\\
\hspace*{12cm} $\Box$

\section{Proof of Theorem 4}

Let $\Lambda\subset \mathbb{R}^n$ be a lattice of rank $n$, the nu-function over $\Lambda$ is defined by
\begin{equation*}
\nu_{\Lambda}(z)=\sum\limits_{x\in \Lambda}z^{|x|_1},
 \end{equation*}
  where $|x|_1=|x_1|+|x_2|+\cdots+|x_n|$ is the $L^1$-norm. Because $z^{|x|_1}$ is not a fixed point under the Fourier transform, the Poisson summation formula for the nut function over lattice $\Lambda$ is unknown.

As mentioned in Section 1, in \cite{008}, we demonstrate that conjecture (1.7) holds true for a lattice associated with a binary code $C$. However, in general cases, the result deviates significantly from conjecture (1.7). In Theorem 4, we present a novel formula for the nut function associated with a ternary code. To establish Theorem 4, we first prove several auxiliary lemmas.

\textbf{Lemma 3.1}\ \ Let $L\subset \mathbb{R}^n$ be any a lattice, $z$ is a given parameter, then $\nu_{3L}(z)=\nu_{L}(z^3)$.

\textbf{Proof:} By the definition of nu-function, we obtain
\begin{equation*}
\nu_{3L}(z)=\sum\limits_{x\in 3L} z^{|x|_1}=\sum\limits_{y\in L} z^{|3y|_1}=\sum\limits_{y\in L} (z^3)^{|y|_1}=\nu_L(z^3).
\end{equation*}
\hspace*{12cm} $\Box$

\textbf{Lemma 3.2}\ \ Let $A_3(C)$ be a lattice associated with the ternary code $C$, then the dual lattice of $A_3(C)$ is given by
\begin{equation*}
A_3(C)^*=\frac{1}{3} A_3(C^{\bot}),
\end{equation*}
where $C^{\bot}$ is the dual code of $C$.

\textbf{Proof:} First, we prove that $A_3(C)^*\subset \frac{1}{3}A_3(C^{\bot})$. For any $\alpha\in A_3(C)^*$, we note that $x\in A_3(C)$ whence $x\in C$, it follows that $<\alpha,x>\in \mathbb{Z}$ for all $x\in C$,  and
\begin{equation*}
<3\alpha,x>\equiv 0\ (\text{mod}\ 3),\quad \forall x\in C.
\end{equation*}
 Thus, $3\alpha\ \text{mod}\ 3\in C^{\bot}$ and $3\alpha\in A_3(C^{\bot})$, which implies that $\alpha \in \frac{1}{3} A_3(C^{\bot})$.

To show that $\frac{1}{3}A_3(C^{\bot}) \subset A_3(C)^*$, for any $\beta\in \frac{1}{3}A_3(C^{\bot})$ or $3\beta\in A_3(C^{\bot})$, we see that $\beta\in A_3(C)^*$. Let $x\in C$, we have $<3\beta\ \text{mod}\ 3,x>=0$, which implies that $<3\beta,x>\equiv 0\ (\text{mod}\ 3)$. Thus, we have $<\beta,x>\in \mathbb{Z}$ for all $x\in C$. Let $y\in A_3(C)$ denote $y\ \text{mod}\ 3=x_0\in C$, then  $<\beta,y>\in \mathbb{Z}$ by $<\beta,x_0>\in \mathbb{Z}$, which leads to $\beta\in A_3(C)^*$ and $\frac{1}{3}A_3(C^{\bot}) \subset A_3(C)^*$. We have lemma 3.2.\\
\hspace*{12cm} $\Box$

\textbf{Lemma 3.3}\ \ Let $C\subset \mathbb{Z}_3^n$ be a ternary code of length $n$, $\alpha$ be any a parameter, then we have
\begin{equation*}
\nu_{A_3(C)}(\tanh\frac{\alpha}{2})=\left(\frac{1+\tanh^3 \frac{\alpha}{2}}{1-\tanh^3 \frac{\alpha}{2}} \right)^n W_C\left(\frac{\tanh \alpha}{2-\tanh\alpha}\right),
\end{equation*}
where $W_C(z)$ is the weight enumerator function.

\textbf{Proof:} For any $x\in C$, we define a set $A_3(x)$ by
\begin{equation*}
A_3(x)=\{y\in \mathbb{Z}^n\ |\ y\equiv x\ (\text{mod}\ 3)\}.
\end{equation*}
It is easy to see that $A_3(x_1) \cap A_3(x_2)=\emptyset$, if $x_1\neq x_2$, so we have
\begin{equation*}
A_3(C)=\mathop{\cup}\limits_{x\in C} A_3(x),\ \text{and}\ \nu_{A_3(C)}(z)=\sum\limits_{x\in C}\nu_{A_3(x)}(z).
\end{equation*}
Let $x=(x_1,x_2,\cdots,x_n)\in C$, then $A_3(x)=(x_1+3\mathbb{Z})\times (x_2+3\mathbb{Z})\times \cdots \times (x_n+3\mathbb{Z})$. Assume $n_0(x)=|\{i|x_i=0\}|$, $n_1(x)=|\{i|x_i=1\}|$ and $n_2(x)=|\{i|x_i=2\}|$. This was not difficult to observe
\begin{equation*}
\nu_{A_3(x)}(z)=\sum\limits_{y\in A_3(x)}z^{|y|_1}=(\sum\limits_{y'\in 3\mathbb{Z}}z^{|y'|})^{n_0(x)} (\sum\limits_{y''\in 3\mathbb{Z}+1}z^{|y''|})^{n_1(x)} (\sum\limits_{y'''\in 3\mathbb{Z}+2}z^{|y'''|})^{n_2(x)}
\end{equation*}
\begin{equation*}
=(\nu_{3\mathbb{Z}}(z))^{n_0(x)} (\nu_{3\mathbb{Z}+1}(z))^{n_1(x)} (\nu_{3\mathbb{Z}+2}(z))^{n_2(x)}.\qquad\qquad\ \ \
\end{equation*}
If the parameter $z$ satisfies $|z|<1$, one has
\begin{equation*}
\nu_{3\mathbb{Z}}(z)=\frac{1+z^3}{1-z^3},\ \text{and}\ \nu_{3\mathbb{Z}+1}(z)=\nu_{3\mathbb{Z}+2}(z)=\frac{z+z^2}{1-z^3}.
\end{equation*}
It follows that
\begin{equation*}
\nu_{A_3(x)}(z)=\left(\frac{1+z^3}{1-z^3}\right)^{n_0(x)} \left(\frac{z+z^2}{1-z^3}\right)^{n_1(x)+n_2(x)}=\left(\frac{1+z^3}{1-z^3}\right)^{n-w(x)} \left(\frac{z+z^2}{1-z^3}\right)^{w(x)}.
\end{equation*}
 From the above calculation, we obtain:
\begin{equation*}
\nu_{A_3(C)}(z)=\sum\limits_{x\in C} \left(\frac{1+z^3}{1-z^3}\right)^{n-w(x)} \left(\frac{z+z^2}{1-z^3}\right)^{w(x)}=\left(\frac{1+z^3}{1-z^3} \right)^n \sum\limits_{x\in C} \left(\frac{z}{1-z+z^2} \right)^{w(x)}.
\end{equation*}
Let $\alpha$ be a parameter such that $z=\tanh \frac{\alpha}{2}$, it is easy to compute that
\begin{equation*}
\frac{z}{1-z+z^2}=\frac{\tanh\alpha}{2-\tanh\alpha}.
\end{equation*}
We have
\begin{equation*}
\nu_{A_3(C)}(\tanh\frac{\alpha}{2})=\left(\frac{1+\tanh^3 \frac{\alpha}{2}}{1-\tanh^3 \frac{\alpha}{2}} \right)^n W_C(\frac{\tanh\alpha}{2-\tanh\alpha}).
\end{equation*}
This is the proof of lemma 3.3.\\
\hspace*{12cm} $\Box$

\textbf{Lemma 3.4}\ \ For any ternary code $C\subset \mathbb{Z}_3^n$, we have
\begin{equation*}
|C|\text{det}(A_3(C))=3^n,
\end{equation*}
where $A_3(C)$ is the lattice associated with $C$, and $\text{det}(A_3(C))$ is the determinant of $A_3(C)$.

\textbf{Proof:} Considering the additive group homomorphism $\sigma: \mathbb{Z}^n \rightarrow \mathbb{Z}_3^n$ given by
\begin{equation*}
(x_1,x_2,\cdots,x_n)\in \mathbb{Z}^n \xrightarrow[]{\ \ \sigma\ \ } (x_1\ \text{mod}\ 3, x_2\ \text{mod}\ 3, \cdots, x_n\ \text{mod}\ 3)\in \mathbb{Z}_3^n.
\end{equation*}
Obviously, $\sigma^{-1}(C)=A_3(C)$, one has the following quotient group isomorphism
\begin{equation*}
\mathbb{Z}^n/A_3(C)\cong \mathbb{Z}_3^n/C.
\end{equation*}
It is well known that
\begin{equation*}
\text{det}(A_3(C))=|\mathbb{Z}^n/A_3(C)|.
\end{equation*}
Thus, we have
\begin{equation*}
|C|\text{det}(A_3(C))=|\mathbb{Z}_3^n|=3^n.
\end{equation*}
\hspace*{12cm} $\Box$

Now, we give a proof of Theorem 4.

\textbf{Proof of Theorem 4:}

Let $C\subset \mathbb{Z}_3^n$ be arbitrary a ternary code of length $n$, $A_3(C)$ be the lattice associated with $C$. By lemma 3.1, we have
\begin{equation*}
\nu_{A_3(C)^*}(\tanh^3 \frac{\beta}{2})=\nu_{3A_3(C)^*}(\tanh \frac{\beta}{2}).  \tag{3.1}
\end{equation*}
By lemma 3.2, we have
\begin{equation*}
\nu_{3A_3(C)^*}(\tanh \frac{\beta}{2})=\nu_{A_3(C^{\bot})}(\tanh \frac{\beta}{2})  \tag{3.2}
\end{equation*}
and by lemma 3.3
\begin{equation*}
\nu_{A_3(C^{\bot})}(\tanh \frac{\beta}{2})=\left(\frac{1+\tanh^3 \frac{\beta}{2}}{1-\tanh^3 \frac{\beta}{2}} \right)^n W_{C^{\bot}}(\frac{\tanh\beta}{2-\tanh\beta}).  \tag{3.3}
\end{equation*}
Combining with (3.1), (3.2) and (3.3), we have
\begin{equation*}
\nu_{A_3(C)^*}(\tanh^3 \frac{\beta}{2})=\left(\frac{1+\tanh^3 \frac{\beta}{2}}{1-\tanh^3 \frac{\beta}{2}} \right)^n W_{C^{\bot}}(\frac{\tanh\beta}{2-\tanh\beta}).  \tag{3.4}
\end{equation*}
By the MacWilliams identity of $C$ (taking $q=3$ in (1.1))
\begin{equation*}
\sum\limits_{c\in C^{\bot}} z^{w(c)}=\frac{1}{|C|}(1+2z)^n \sum\limits_{c\in C} (\frac{1-z}{1+2z})^{w(c)}.  \tag{3.5}
\end{equation*}
Let $z=\tanh\beta/(2-\tanh\beta)$, based on the relation $e^{-2\beta}=3\tanh\alpha/(8-5\tanh\alpha)$ between $\alpha$ and $\beta$, we get
\begin{equation*}
\frac{1-z}{1+2z}=\frac{2-2\tanh\beta}{2+\tanh\beta}=\frac{4}{3e^{2\beta}+1}=\frac{\tanh\alpha}{2-\tanh\alpha},  \tag{3.6}
\end{equation*}
According to (3.5) and (3.6), it follows that
\begin{equation*}
W_{C^{\bot}}(\frac{\tanh\beta}{2-\tanh\beta})=\frac{1}{|C|}\left(\frac{2+\tanh\beta}{2-\tanh\beta} \right)^n W_C(\frac{\tanh\alpha}{2-\tanh\alpha}).  \tag{3.7}
\end{equation*}
In (3.3) we replace $C^{\bot}$ by $C$ and change $\beta$ to $\alpha$, which gives
\begin{equation*}
\nu_{A_3(C)}(\tanh \frac{\alpha}{2})=\left(\frac{1+\tanh^3 \frac{\alpha}{2}}{1-\tanh^3 \frac{\alpha}{2}} \right)^n W_{C}(\frac{\tanh\alpha}{2-\tanh\alpha}),  \tag{3.8}
\end{equation*}
Based on (3.4), (3.7), (3.8), and note that
\begin{equation*}
\frac{2+\tanh\beta}{2-\tanh\beta}\cdot \frac{1-\tanh\frac{\beta}{2}+\tanh^2 \frac{\beta}{2}}{1+\tanh\frac{\beta}{2}+\tanh^2 \frac{\beta}{2}}=1,
\end{equation*}
one can get
\begin{equation*}
\nu_{A_3(C)^*}(\tanh^3 \frac{\beta}{2})=\frac{1}{|C|}\left(\frac{1+\tanh \frac{\beta}{2}}{1-\tanh \frac{\beta}{2}} \right)^n \left(\frac{1-\tanh^3 \frac{\alpha}{2}}{1+\tanh^3 \frac{\alpha}{2}} \right)^n \nu_{A_3(C)}(\tanh \frac{\alpha}{2}).  \tag{3.9}
\end{equation*}
Finally, by lemma 3.4 and take $\Lambda=A_3(C)$, (3.9) is equivalent to
\begin{equation*}
\nu_{\Lambda^*}(\tanh^3 \frac{\beta}{2})=\frac{\text{det}(\Lambda)}{3^n} \left[\frac{(1+\tanh\frac{\beta}{2})(1-\tanh^3 \frac{\alpha}{2})}{(1-\tanh\frac{\beta}{2})(1+\tanh^3 \frac{\alpha}{2})} \right]^n \nu_{\Lambda}(\tanh \frac{\alpha}{2}).
\end{equation*}
This is the whole proof of Theorem 4.\\
\hspace*{12cm} $\Box$

\section{Counter-examples for the Conjecture in General Cases}

In this section, we present counter-examples to demonstrate that the original conjecture 1 in \cite{20} does not hold in general cases.

\textbf{Counter-Example:}\ \ Let $n=1$, $k=5$, $C=\{0\}$, $\Lambda=A_5(C)=5\mathbb{Z}$, then $\Lambda^*=\frac{1}{5}\mathbb{Z}$, $\alpha$ and $\beta$ be the parameters such that $e^{-2\beta}=\tanh\alpha$, we have
\begin{equation*}
2^{\frac{1}{2}}\nu_{\Lambda^*}(\tanh^2 \frac{\beta}{2})\neq\text{det}(\Lambda)(\sinh(2\beta))^{\frac{1}{2}}\nu_{\Lambda}(\tanh\frac{\alpha}{2}).  \tag{4.1}
\end{equation*}

\textbf{Proof of Counter-Example:}\ \ Since $\nu_{\mathbb{Z}}(z)=\frac{1+z}{1-z}$ holds for $0<|z|<1$, note that $0<|\tanh\frac{\alpha}{2}|<1$, $0<\tanh^2 \frac{\beta}{2}<1$, we have
\begin{equation*}
\nu_{\Lambda}(\tanh\frac{\alpha}{2})=\nu_{5\mathbb{Z}}(\tanh\frac{\alpha}{2})=\nu_{\mathbb{Z}}(\tanh^5 \frac{\alpha}{2})=\frac{1+\tanh^5 \frac{\alpha}{2}}{1-\tanh^5 \frac{\alpha}{2}}
\end{equation*}
\begin{equation*}
=\frac{1+(\frac{e^{\alpha}-1}{e^{\alpha}+1})^5}{1-(\frac{e^{\alpha}-1}{e^{\alpha}+1})^5}=\frac{e^{5\alpha}+10e^{3\alpha}+5e^{\alpha}}{5e^{4\alpha}+10e^{2\alpha}+1},
\end{equation*}
and
\begin{equation*}
\nu_{\Lambda^*}(\tanh^2 \frac{\beta}{2})=\nu_{\frac{1}{5}\mathbb{Z}}(\tanh^2 \frac{\beta}{2})=\nu_{\mathbb{Z}}(\tanh^{\frac{2}{5}} \frac{\beta}{2})=\frac{1+\tanh^{\frac{2}{5}} \frac{\beta}{2}}{1-\tanh^{\frac{2}{5}} \frac{\beta}{2}}.
\end{equation*}
In order to prove (4.1), we only need to show that
\begin{equation*}
2^{\frac{1}{2}} \frac{1+\tanh^{\frac{2}{5}} \frac{\beta}{2}}{1-\tanh^{\frac{2}{5}} \frac{\beta}{2}}\neq 5(\sinh 2\beta)^{\frac{1}{2}} \frac{e^{5\alpha}+10e^{3\alpha}+5e^{\alpha}}{5e^{4\alpha}+10e^{2\alpha}+1},
\end{equation*}
which is equivalent to
\begin{equation*}
\frac{1+\tanh^{\frac{1}{2}} \frac{\beta}{2}}{1-\tanh^{\frac{1}{2}} \frac{\beta}{2}}\neq \frac{5}{2} (e^{2\beta}-e^{-2\beta})^{\frac{1}{2}} \frac{e^{5\alpha}+10e^{3\alpha}+5e^{\alpha}}{5e^{4\alpha}+10e^{2\alpha}+1}.  \tag{4.2}
\end{equation*}
The validity of inequality (4.2) can be demonstrated by selecting two specific values for $\alpha$ and $\beta$. For example, when $\alpha=1$, solving the equation $e^{-2\beta}=\tanh\alpha$ yields $\beta\approx 0.136$. This verification concludes the proof of the counterexample.\\

\hspace*{12cm} $\Box$

In this counterexample, we employ a modulo of $k=5$ and construct the lattice $\Lambda=A_5(C)$ using the zero code $C=\{0\}$. It is worth noting that numerous counterexamples to the original conjecture exist in general cases. For any modulo $k\geqslant 3$, it is possible to assign a value of one. In our approach, we set $\beta=1$ and derive $\alpha\approx 0.136$ using the relationship $e^{-2\alpha}=\tanh\beta$. We then apply the lattice $\Lambda=A_k(C)$, constructed using the zero code $C=\{0\}$, for all $k\geqslant 3$. To illustrate that the conjecture does not hold for general lattices, Table 1 presents the results for both sides of conjecture 1 \cite{20}, covering modulo values from $k=3$ to 10.

\begin{table}[htbp]
\centering
\caption{\small{The results of both sides in conjecture 1 from modulo $k=3$ to 10}}
\begin{tabular}{ccc}
\hline
$k$ & the left side & the right side\\
\hline
3 & 5.6169 & 5.7169\\
4 & 7.4189 & 7.6181\\
5 & 9.2328 & 9.5222\\
6 & 11.0528 & 11.4266\\
7 & 12.8762 & 13.3310\\
8 & 14.7017 & 15.2355\\
9 & 16.5287 & 17.1399\\
10 & 18.3567 & 19.0443\\
\hline
\end{tabular}
\end{table}

\section*{Acknowledgments}

This work was supported by Information Security School-Enterprise Joint Laboratory (Dongguan Institute for Advanced Study, Greater Bay Area) and Project (No. H24120002).

\end{document}